\documentclass[aps,prl,balance,groupedaddress,amsmath,reprint,10pt,twocolumn]{revtex4-1}
\usepackage[latin1]{inputenc}    
\usepackage{graphicx}   
\usepackage{xspace}
\usepackage{rotating}
\usepackage{caption}
\usepackage{color}
\usepackage{gensymb}
\usepackage{amsfonts}
\usepackage{changepage}
\usepackage{graphicx}
\usepackage{float}
\usepackage{amsmath}
\usepackage{amssymb}
\usepackage{tabu}
\usepackage{appendix}
\usepackage{amsmath,amssymb}
\usepackage{tikz}
\usepackage{color}
\setcitestyle{super}
\usepackage{lineno}
\usepackage[labelformat=empty]{caption}
\usepackage{hyperref}
\hypersetup{%
        colorlinks=true,
        linkcolor=blue,
        citecolor=blue,
        urlcolor=blue,
      }

\usepackage[euler]{textgreek}
\usepackage[ugly]{units}
\usepackage{xcolor}
\usepackage{ wasysym }

\begin{document}

\title{Circuit quantum acoustodynamics with surface acoustic waves} 
\author{R.~Manenti$^{1}$}   
\author{A.~F.~Kockum$^2$}
\author{A.~Patterson$^{1}$}
\author{T.~Behrle$^{1}$} 
\author{J.~Rahamim$^{1}$}
\author{G.~Tancredi$^{1}$}
\author{F.~Nori$^{2,3}$}
\author{P.~J.~Leek$^1$}
\affiliation{$^1$Clarendon Laboratory, Department of Physics, University of Oxford, OX1 3PU, Oxford, United Kingdom}
\affiliation{$^2$Center for Emergent Matter Science, RIKEN, Saitama 351-0198, Japan}
\affiliation{$^3$Physics Department, The University of Michigan, Ann Arbor, Michigan 48109-1040, USA}

\date{\today}

\pacs{85.25.Qc}

\keywords{quantum acoustics; surface acoustic waves; superconducting devices;} 
\maketitle
\textbf{The experimental investigation of quantum devices incorporating mechanical resonators has opened up new frontiers in the study of quantum mechanics at a macroscopic level \cite{Poot:2012,Aspelmeyer:2014}. Superconducting microwave circuits have proven to be a powerful platform for the realisation of such quantum devices, both in cavity optomechanics\cite{Teufel:2011,LaHaye:2009}, and circuit quantum electrodynamics (QED)\cite{O'Connell:2010,Pirkkalainen:2015}. While most experiments to date have involved localised nanomechanical resonators, it has recently been shown that propagating surface acoustic waves (SAWs) can be piezoelectrically coupled to superconducting qubits \cite{Gustafsson:2014,Aref:2016}, and confined in high-quality Fabry-Perot cavities up to microwave frequencies in the quantum regime\cite{Manenti:2016}, indicating the possibility of realising coherent exchange of quantum information between the two systems. Here we present measurements of a device in which a superconducting qubit is embedded in, and interacts with, the acoustic field of a Fabry-Perot SAW cavity on quartz, realising a surface acoustic version of cavity quantum electrodynamics. This quantum acoustodynamics (QAD) architecture may be used to develop new quantum acoustic devices in which quantum information is stored in trapped on-chip surface acoustic wavepackets, and manipulated in ways that are impossible with purely electromagnetic signals, due to the $10^5$ times slower speed of travel of the mechanical waves.}

The study of the quantum nature of mechanical systems has rapidly increased in the last decade\cite{Poot:2012,Aspelmeyer:2014,Treutlein:2014}. The primary goal of these experiments has been the demonstration of the quantum behaviour of macroscopic objects when suitably isolated from their environment, with the intent to corroborate the validity of quantum mechanics at macroscopic scales. Pioneering work has now experimentally proved the possibility to prepare mechanical objects close to their quantum ground state\cite{Teufel:2011, Chan:2011} and to coherently manipulate their state\cite{O'Connell:2010}. These results have encouraged new lines of research utilising mechanical quantum devices, including the development of microwave-optical converters\cite{Andrews:2014}, mechanical quantum memories \cite{Palomaki:2013} and quantum limited amplifiers \cite{Massel:2011}, the generation of squeezed vacuum states of mechanical objects \cite{Wollman:2015, Pirkkalainen:2015}, and the detection of non-classical correlations of photon-phonon pairs \cite{Riedinger:2016}.

A highly successful architecture for the exchange of single quanta between coupled quantum systems is the solid-state version of cavity quantum electrodynamics (QED), known as circuit QED\cite{Wallraff:2004}, in which the electrical interaction between a qubit and a high-quality microwave resonator offers the possibility to reliably control, store, and read out quantum bits of information on a chip.  Although many quantum experiments involving mechanical objects have been reported that use an optomechanical coupling between an electric field and a mechanical system \cite{Aspelmeyer:2014}, a parallel series of investigations have employed such a circuit-QED type of interaction between mechanical resonators and superconducting qubits \cite{LaHaye:2009, O'Connell:2010, Pirkkalainen:2013, Rouxinol:2016}, which in principle enables full control of the quantum state of the mechanical mode via the qubit.

In contrast to these experiments involving localised mechanical modes, studies have also recently emerged on the coupling of superconducting qubits to traveling surface acoustic waves (SAWs)\cite{Gustafsson:2014,Aref:2016}, which are mechanical perturbations that propagate on the surface of a crystal\cite{Morgan:2007}, and are naturally coupled to superconducting circuits using the piezoelectric effect. As well as being of fundamental interest to study such acoustic waves at the quantum level, they may find uses in quantum signal processing, since their slow speed of travel (five orders of magnitude slower than light) means many-wavelength signals can be manipulated on a mm-scale chip \cite{Aref:2016,Kockum:2014}. It has been demonstrated that Fabry-Perot SAW cavities formed using superconducting surface Bragg mirrors can reach quality factors in the $10^5$ range at microwave frequencies\cite{Manenti:2016, Magnusson:2015}, opening up the possibility of realising surface acoustic cavity QED, either with superconducting qubits, or with other solid-state quantum systems\cite{Schuetz:2015}. SAW cavities have also been proposed as a potential quantum acousto-optic transducer between superconducting qubits and optical photons by exploiting stimulated Brillouin scattering\cite{Shumeiko:2016}.

In this work, we present measurements of a device in which a tuneable transmon qubit\cite{Koch:2007} is piezoelectrically coupled to a surface acoustic wave cavity, displaying a surface acoustic version of cavity QED which we call circuit quantum acoustodynamics (QAD). We characterise the dispersive interaction between the two systems in several ways. First, we measure the frequency shift of the acoustic mode as the qubit is flux tuned. Secondly, we measure the acoustic Stark shift of the qubit due to the population of the mechanical resonator and we observe a preferential coupling of the qubit to one longitudinal mode of the acoustic cavity. We extract the coupling and we show that it is in agreement with theoretical expectations. In order to demonstrate the possibility to control the device in the time domain, we show a time delayed Stark shift made possible by the slow travel of the wave. We also present spectroscopic measurements of the qubit via the Stark shift of the acoustic cavity, indicating that SAWRs can in principle be adopted as an alternative qubit readout scheme in quantum information processors.

\begin{figure}[t]
\begin{center}
\includegraphics[width=0.95\linewidth]{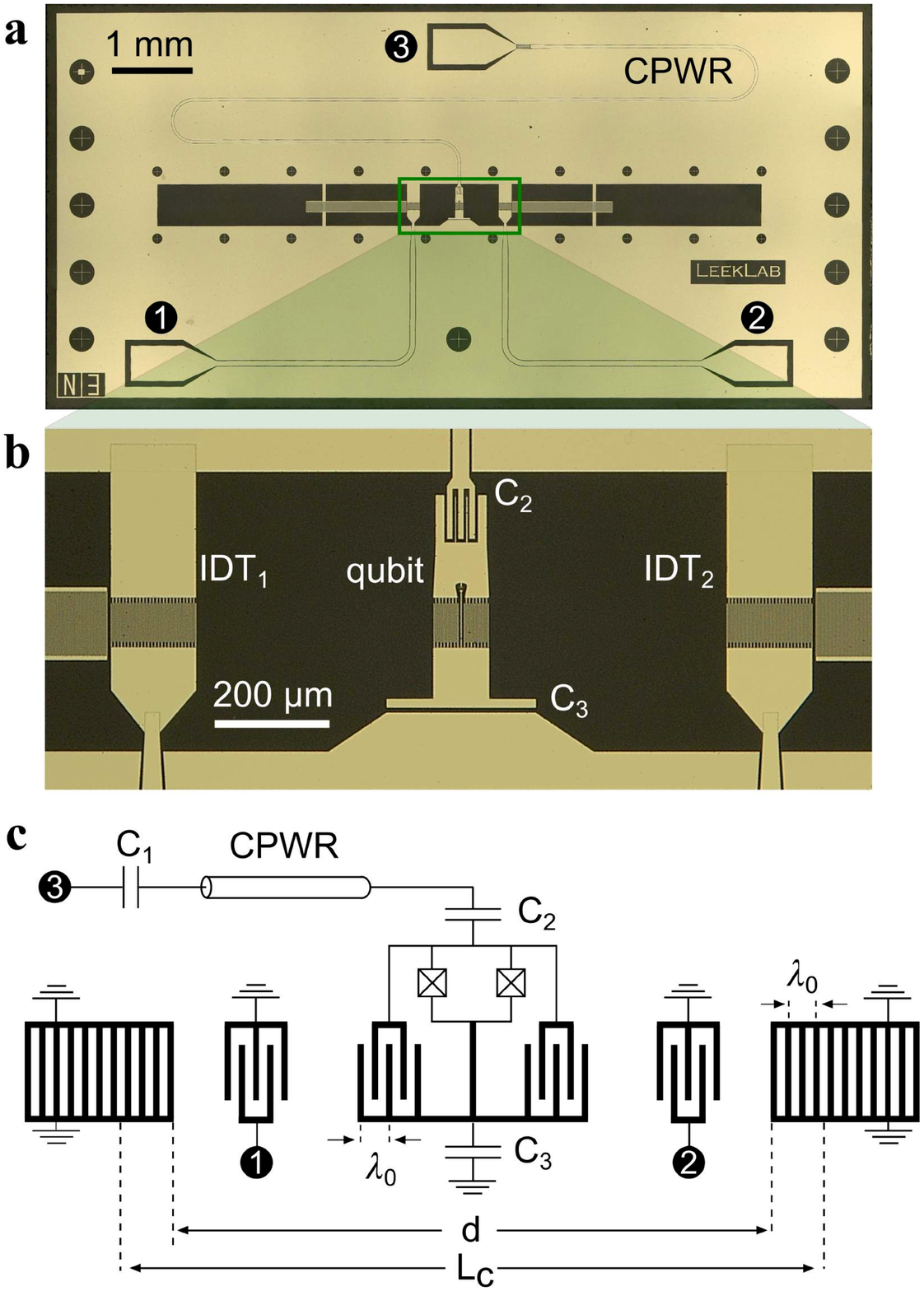}
\caption{
$\textbf{\!\! Figure 1 $|$ Circuit quantum acoustodynamics device.}$ \textbf{a}, 
Optical image of the measured device. In the centre of the chip, a transmon is embedded in a SAW cavity. A coplanar waveguide resonator (CPWR) is coupled to the transmon and measured via port 3. The SAW cavity is probed via two interdigitated transducers (IDTs) connected to ports 1 and 2. \textbf{b}, Close-up image showing the transmon qubit and SAW IDTs in between the two Bragg gratings that form the SAW cavity. \textbf{c}, Equivalent electrical circuit of the device incorporating a spatial schematic of the SAW cavity. The geometrical parameters $\lambda_0$, $d$ and $L_\textrm{c}$ denote the wavelength, the distance between the two Bragg gratings, and the effective length of the cavity, respectively.
\label{fig:1} }
\end{center}
\end{figure}

Our measured device (see Fig.~\ref{fig:1}) is fabricated on ST-X quartz, on which the free SAW speed is $v_\textrm{f}\approx \unit[3158]{m/s}$ at room temperature\cite{Morgan:2007}. This traveling mode is excited by applying an oscillating voltage to the electrodes of an interdigitated transducer (IDT) patterned on the surface of the substrate. The propagating SAW is confined to a small region of the chip between two Bragg mirrors facing each other forming a Fabry-Perot acoustic cavity; each mirror consists of a regular array of shorted metallic strips. A tuneable transmon qubit is situated in the middle of the SAW cavity and consists of a SQUID shunted by an interdigitated capacitance with periodicity $\lambda_0$ matching the SAW IDTs\cite{Sup}. The transmon is also coupled to an auxiliary coplanar waveguide resonator (CPWR) employed for independent dispersive qubit readout \cite{Wallraff:2005}. All measurements presented hereafter have been performed at the base temperature  $T\approx\unit[10]{mK}$ of a dilution refrigerator. Microwave ports 1 and 2 are connected to room temperature via low ($\approx 16~\rm{dB}$) attenuation lines in order to easily populate and measure the SAW cavity modes, while port 3 is highly attenuated ($\approx 70~\rm{dB}$) such that the CPWR and qubit are close to their quantum-mechanical ground states.

\begin{figure}
\centering
\includegraphics[width=0.95\linewidth]{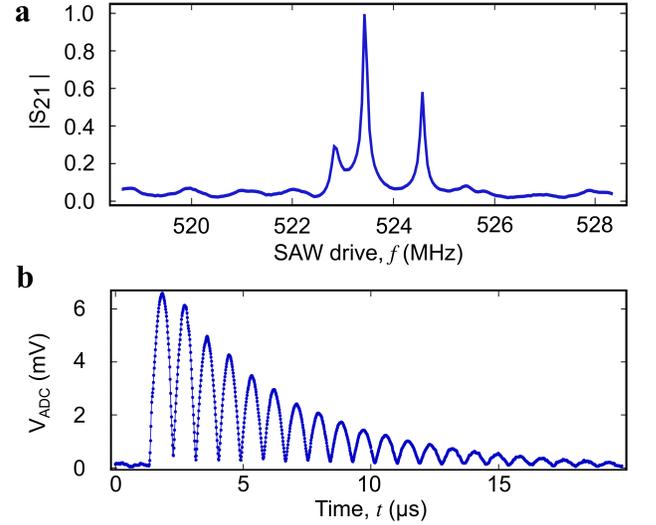}
\caption{ $\textbf{Figure 2 $|$ SAW cavity response.}$ \textbf{a}, Normalised linear magnitude of the measured transmission coefficient $S_{21}$ of the 2-port SAWR (blue solid line). The transmitted signal has been acquired with a vector network analyser with input power set at $\unit[-30]{dBm}$. \textbf{b}, Time resolved measurement of the 2-port SAWR. This measurement has been performed by applying a $\unit[800]{ns} \lesssim 2L_\textrm{c}/v_\textrm{e}$ electrical pulse to $\textrm{IDT}_1$ and acquiring the output signal from $\textrm{IDT}_2$. The graph shows the voltage difference at the input of the acquisition card.}
 \label{fig:2}
\end{figure}

The SAW cavity contains two transducers for the excitation and detection of the acoustic wave. Figure~\ref{fig:2}\textcolor{blue}{a}  shows the transmission coefficient $S_\textrm{21}$ of the cavity, measured via the two IDTs. The frequency of the central mechanical mode is $f_{\textrm{m2}}=\unit[523.435]{MHz}$, while side peaks seen at  $f_{\textrm{m1}}=\unit[522.83]{MHz}$ and $f_{\textrm{m3}}=\unit[524.58]{MHz}$ are likely to be additional mechanical modes. The quality factors of these modes, $Q_\textrm{m1,m2,m3}=\{4820,6980,7580\}$, are obtained from additional measurements\cite{Sup} of $S_{11}$. Since the periodicity of the IDTs is set to $\lambda_0=\unit[6]{\mu m}$ in fabrication, the central mode frequency is consistent with a speed of sound of $v_\textrm{e}=f_\textrm{m2}\lambda_0=\unit[3140.6]{m/s}$, assuming a symmetric device. The slight difference between $v_\textrm{e}$ and the textbook room temperature value $v_\textrm{f}$ may be due to slight device asymmetry and/or stiffness tensor changes or crystal contraction at millikelvin temperatures. Note that the SAW cavity modes are not in their ground state, due to the mode frequencies $f_{\textrm{m}i}\lesssim k_BT/h$ and low-attenuation connections to room temperature.

The acoustic nature of the observed resonant modes can be further tested by measuring the device response to a short coherent drive pulse (see Fig.~\ref{fig:2}\textcolor{blue}{b}). An exponentially decaying train of pulses is clearly observed in the response, consistent 
 with the applied $800~\rm{ns}$ pulse reflecting back and forth between the mirrors of the cavity. The decay time of the pulses, the lifetime of phonons in the SAW cavity, is $\tau\approx\unit[2.4]{\mu s}$. 
The pulses measured in the response are separated by $\Delta t = \unit[870]{ns}$, consistent 
with a cavity length of $L_\textrm{c} =v_\textrm{e} \Delta t /2 \approx \unit[1365]{\mu m}$. As expected, the cavity length is slightly longer than the distance between the two gratings $d=\unit[1260]{\mu m}$. This is consistent 
with the fact that SAWs slightly penetrate into the gratings by an amount $L_\textrm{p} = (L_\textrm{c}-d)/2=\unit[55]{\mu m} $ before being efficiently reflected. 
The frequency spacing of adjacent modes (the free spectral range) of the cavity are also consistent 
with the same cavity length, $L_\textrm{c} = v_\textrm{e}/2\left|f_{\textrm{m2}}-f_{\textrm{m3}}\right| \approx \unit[1365]{\mu m}$, where we have used the two higher-quality modes $f_{\textrm{m2}}$ and $f_{\textrm{m3}}$. Note however that $f_\textrm{m1}$ is slightly closer in frequency to $f_\textrm{m2}$. This asymmetric behaviour in the frequency domain may be due to the fact that the grating stopband does not perfectly coincide with the resonant frequency of the IDTs\cite{Uno:1982}. 

Having characterised the SAW cavity, we now proceed to examine its interaction with the superconducting qubit, a flux-tuneable transmon. An appropriate quantum-mechanical description of this system (including the readout CPWR) is the generalized Jaynes-Cummings Hamiltonian of two resonators both coupled to the same transmon qubit (for reasons that will become clear, we will only consider the central acoustic mode at $f_\textrm{m2}$, which we hereafter refer to as the SAW resonator, SAWR):
\begin{eqnarray}
\!\!\!\hat{H} /h& = & 
\sum_{j}f_{j}(\Phi)|j\rangle\langle j|+f_\textrm{r}\hat{a}^{\dagger}\hat{a}+f_\textrm{m2}\hat{b}^{\dagger}\hat{b}+\nonumber \\
&  &\!\!\!\!\!\!\!\!\!\!\!\!\!\!\!\!\!\!\!\!\!\!\!+\sum_{i,j}  \big[g_{ij}(\Phi) |i\rangle\langle j|\left(\hat{a}+\hat{a}^{\dagger}\right)+\lambda_{\textrm{m2},ij}(\Phi)|i\rangle\langle j| (\hat{b}+\hat{b}^{\dagger})\big],
\label{eq:hamiltonian}
\end{eqnarray}
where $f_j (\Phi) $ are the flux dependent transmon transition frequencies, $f_\textrm{r}=\unit[5.83]{GHz}$ is the CPWR frequency, $\hat{a}$ ($\hat{a}^\dagger$) and $\hat{b}$ ($\hat{b}^\dagger$) are the annihilation (creation) operators of the microwave cavity and of the mechanical resonator, respectively, and $g_{ij}(\Phi)$ [$\lambda_{\textrm{m2},ij}(\Phi)$] is the coupling strength between the qubit and the CPWR (SAWR). Hereafter, we will denote the coupling strength between the CPWR (SAWR) and the first energy level of the qubit simply by $g$ ($\lambda_{\textrm{m2}}$). The transition frequency $f_\textrm{q}$ between the first two energy levels of the qubit is tuned by applying an external magnetic flux $\Phi$ to its superconducting loop, and its value is given by:
\begin{equation}
h f_\textrm{q}(\Phi)=\sqrt{8E_{\textrm{C}}E_{\textrm{J}0}\cos\left|\pi \Phi / \Phi_0 \right|}-E_{\textrm{C}},\label{eq:qubitfrequency}
\end{equation}
where $\Phi_0=h/2e$ is the magnetic flux quantum, and $E_{\textrm{C}}=h\times\unit[0.31]{GHz}$, $ E_{\textrm{J0}} =h\times\unit[10.7]{GHz}$ are the Coulomb and maximum Josephson energy of the qubit (obtained from standard qubit spectroscopy). The electrical coupling  between the CPWR and the qubit mainly originates from the capacitance $C_\textrm{2}$ shown in Fig.~\ref{fig:1}\textcolor{blue}{b}. The acoustic coupling is instead due to the potential difference generated by the acoustic wave on the electrodes of the qubit and is given by:
\begin{equation}
\lambda_\textrm{m2}(\Phi,f)=\frac{e}{h} \frac{C_\textrm{q}}{C_\Sigma}   \left(\frac{E_\textrm{J}(\Phi)}{2E_\textrm{C}}\right)^{1/4} \frac{e_\textrm{pz}}{\varepsilon} \sqrt{ \frac{\hbar}{2  \rho A_\textrm{c} v_\textrm{e}}}A(f)  , \label{eq:couplingstrength}
\end{equation}
where $e$ is the electron charge, $C_\textrm{q}$ and $C_\Sigma$ are the qubit capacitance and the total capacitance seen by the qubit respectively, $e_\textrm{pz}$ is the piezoelectric coupling coefficient, $\varepsilon$ is the substrate permittivity, $\rho$ is the substrate mass density, $A_\textrm{c}$ is the acoustic cavity area, and $A(f)$ is a normalized array factor\cite{Sup}. From values related to our experiment, this formula predicts $\lambda_\textrm{m2} = \unit[6.0]{MHz}$ (for a qubit frequency of $f_\textrm{q}=\unit[2.52]{GHz}$, for comparison with later measurements). 

\begin{figure}[t]
\begin{center}
\includegraphics[width=0.95\linewidth]{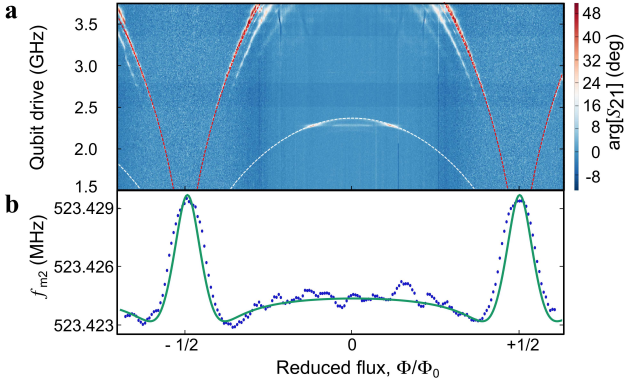}
\end{center}
\caption{
\label{fig:3} $\textbf{\!\!Figure 3 $|$ Flux dependent acoustic shift.}$ \textbf{a}, Qubit spectroscopy performed with the CPWR as a function of reduced magnetic flux. The red dashed line is a fit to equation~\eqref{eq:qubitfrequency} and indicates the qubit transition frequency $f_\textrm{q}(\Phi)$. The white dashed line, with analytical form $f_\textrm{q}(\Phi)/2$, denotes an excitation of the qubit via a two photon process. \textbf{b}, 
Measured resonant frequency of the acoustic mode $f_\textrm{m2}$ as a function of reduced flux at fixed SAW drive power of $\unit[-80]{dBm}$ (blue points) and numerical model based on equation~\eqref{eq:hamiltonian} (green solid curve).}
\end{figure}

As a first probe of the interaction between the qubit and the SAW mode, we measure both the frequency of the qubit (via the CPWR) and the acoustic mode $f_\textrm{m2}$ as a function of magnetic flux $\Phi$ (see Fig.~\ref{fig:3}).
The qubit frequency (see Fig.~\ref{fig:3}\textcolor{blue}{a}) fits well to equation~\eqref{eq:qubitfrequency}, while the SAW mode frequency (see Fig.~\ref{fig:3}\textcolor{blue}{b}) also shows a flux dependence with the same periodicity.
By fitting the experimental curve of Fig.~\ref{fig:3}\textcolor{blue}{b} with a QuTiP numerical model\cite{Johansson:2013} based on equation~\eqref{eq:hamiltonian}, we can extract the value of the acoustic coupling and we find $\lambda_\textrm{m2}=\unit[ 5.7 \pm 0.5 ]{MHz}$ at $f_\textrm{q} = \unit[2.52]{GHz}$. In this model, the transmon coupling strengths and level spacings are calculated by diagonalising the Hamiltonian for a Cooper-pair box including many charge states\cite{Koch:2007}.
The free parameters of the model are the coupling strength $\lambda_\textrm{m2}$, the asymmetry of the critical currents of the two Josephson junctions [$d_\textrm{sym}=(I_{c1}-I_{c2})/(I_{c1}+I_{c2})=0.09$] and the effective temperature of the device ($T=\unit[85]{mK}$).
The additional SAW modes at $f_\textrm{m1}$ and $f_\textrm{m3}$ do not show any detectable flux dependence. This is in agreement with the expectation that these modes are antisymmetric with respect to the center of the cavity, whereas the central mode $f_\textrm{m2}$ and the qubit transducer geometry are both symmetric.

\begin{figure}[t]
\begin{center}
\includegraphics[width=0.95\linewidth]{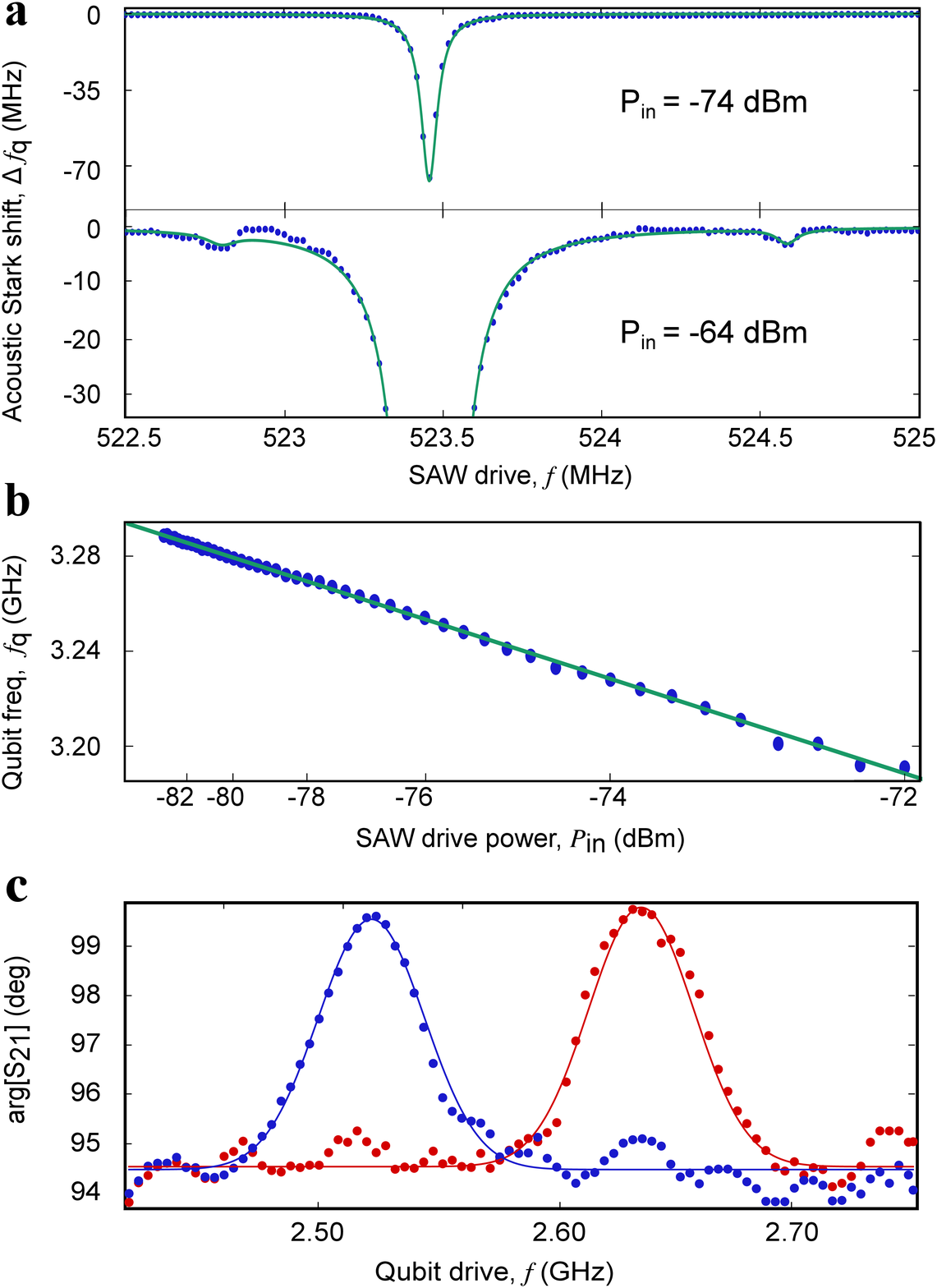}
\caption{
\label{fig:4} $\textbf{\!\!Figure 4 $|$ Acoustic Stark shift.}$ \textbf{a}, Qubit frequency as a function of SAW drive at two different powers: $\unit[-74]{dBm}$ (top panel) and $\unit[-64]{dBm}$ (bottom panel). The green cruves are Lorentzian fits to the data points.  \textbf{b}, Qubit frequency as a function of SAW drive power $P_\textrm{in}$ at a fixed SAW frequency $f_\textrm{m2}$ (blue points). The green solid line is a linear fit to the data.
\textbf{c}, Qubit spectroscopy performed with the SAWR at two different flux values [0.395$\Phi_0$ (red points) and 0.403$\Phi_0$ (blue points)]. The solid lines are gaussian fits to the data points.
}
\end{center}
\end{figure}

A second method to investigate the acoustic coupling between the qubit and the SAW is to measure the AC Stark shift between the two systems in the far-detuned limit (see Fig.~\ref{fig:4}), in which the Hamiltonian of the SAW-qubit system becomes: 
\begin{equation}
\hat{H}_{\textrm{disp}}/h\approx f_\textrm{m2} \hat{b}^\dagger \hat{b} + \frac{1}{2} \left(f_\textrm{q} + 2\chi \hat{b}^\dagger \hat{b} + \chi\right) \hat{\sigma}_z,
\end{equation}
where $\hat{\sigma}_z$ is a Pauli operator and we have approximated the transmon as a two-level system for simplicity. We first set the magnetic flux such that the qubit frequency is $f_\textrm{q}=3.29~\rm{GHz}$, and the qubit-SAW detuning is $\Delta = 2.77~\rm{GHz}\gg\lambda_\textrm{m2}$. In Fig.~\ref{fig:4}\textcolor{blue}{a}, we show the qubit frequency shift as a function of SAW drive frequency close to the acoustic modes at two different drive powers. At the lower power of $P_\textrm{in}=-74~\textrm{dBm}$, we clearly observe a qubit frequency shift only at the frequency of the central SAW mode $f_\textrm{m2}$.  The shift fits well to a Lorentzian centered at $f_\textrm{m2}$, and has a FWHM of $\unit[60\pm5]{kHz}$, close to that obtained for the SAW mode measured via $S_\textrm{21}$. No such frequency shift is observed at the other SAW mode frequencies $f_\textrm{m1}$ and $f_\textrm{m3}$ until higher power. From a second measurement at $P_\textrm{in}=-64~\textrm{dBm}$, high enough to observe small shifts at $f_\textrm{m1}$ and $f_\textrm{m3}$, we can estimate the coupling of the qubit to these additional modes. Assuming that $\lambda_\textrm{m2}=\unit[5.7]{MHz}$ from the fit to the flux dependence of $f_\textrm{m2}$, and taking into account the different drive powers of the two experiments, we can estimate the coupling to the two side modes to be $\lambda_\textrm{m1} \approx \unit[380]{kHz}$ and $\lambda_\textrm{m3} \approx \unit[340]{kHz}$, more than an order of magnitude lower than the coupling to the central mode.
Figure~\ref{fig:4}\textcolor{blue}{b} illustrates the qubit frequency shift as a function of drive power at frequency $f_\textrm{m2}$. The shift is observed to be linear, in agreement with the AC Stark effect.

\begin{figure}[t]
\begin{center}
\includegraphics[width=0.95\linewidth]{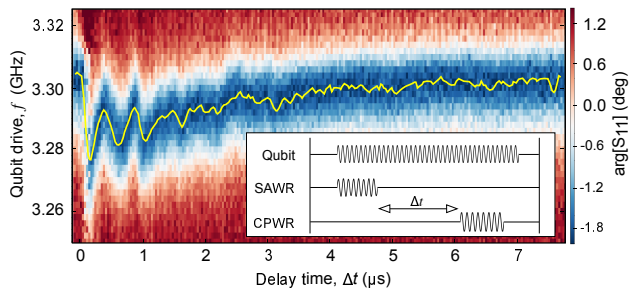}
\caption{\label{fig:5} $\textbf{\!\!Figure 5 $|$ Time-delayed acoustic Stark shift.}$ 
Measured time-delayed acoustic Stark shift of the qubit. The yellow solid line indicates the qubit frequency. Inset: pulse scheme related to this experiment: a continuous drive excites the qubit, while two short \unit[100]{ns} pulses delayed by $\Delta t$ drive the SAWR and the CPWR for the readout.}
\end{center}
\end{figure}

In a complementary experiment, we measure the Stark shift of the SAW mode frequency when the qubit excited state is populated (see Fig.~\ref{fig:4}\textcolor{blue}{c}), an acoustic equivalent of circuit QED dispersive qubit readout\cite{Wallraff:2005}. This measurement can be used to extract an independent estimate of the acoustic coupling $\lambda_\textrm{m2}$. We carry out such a measurement at two different values of the magnetic flux, in both cases measuring the phase shift $\Delta\phi$ of a probe drive at $f_\textrm{m2}$ under strong driving of the qubit. We have $\Delta \phi = 2 \arctan \left(|\chi|/\kappa_\textrm{m2}\right)$, where $\kappa_\textrm{m2}$ is the mechanical mode linewidth, $\chi = -\lambda^2_\textrm{m2} E_\textrm{C}/\Delta (\Delta - E_\textrm{C})$ is the dispersive shift for a transmon and $\Delta = f_\textrm{q} -f_\textrm{m2} $ is the frequency detuning.  From Fig.~\ref{fig:4}\textcolor{blue}{c}, we have $ \Delta = \unit[2.00]{GHz}$ and $\Delta\phi= 4.9^{\circ} \pm 0.3^{\circ}$ and since $\kappa_\textrm{m2} = \unit[75]{kHz}$, we find $|\chi|=3.2~\rm{kHz}$ and $\lambda_\textrm{m2} = \unit[5.9\pm 0.2]{MHz}$, in close agreement with our earlier estimate obtained by fitting the flux dependence of $f_\textrm{m2}$. We can now use this measurement of $\chi$ to obtain an estimate for the average coherent phonon population $\bar{n}=\langle \hat{b}^\dagger \hat{b}\rangle$ of the SAW mode in the experiment shown in Fig.~\ref{fig:4}\textcolor{blue}{a}, obtaining $\bar{n}\approx 10^4 (10^5)$ for $P_\textrm{in}=\unit[-74 (-64)]{dBm}$. These values are within the limit in which the dispersive Hamiltonian remains valid\cite{Boissonneault:2009}, $\lambda_\textrm{m2} \sqrt{\bar{n}} < \Delta$.

We finally report an experiment in which we use the slow travel of the acoustic wave to apply a time-delayed Stark shift to the qubit, occurring as a SAW pulse passes the qubit inside the SAW cavity (see Fig.~\ref{fig:5}). A short $\unit[100]{ns}$ pulse is first applied to one SAW IDT, then a time delayed pulsed measurement of the qubit is subsequently carried out via the CPWR (with a $100~\rm{ns}$ measurement pulse). A continuous drive is applied to the qubit throughout the experiment, the frequency of which is varied to determine the qubit frequency. The qubit is observed to shift lower in frequency initially at a time $170~\rm{ns}$ after the SAW pulse is applied, exactly consistent with the time-of-flight of the SAW pulse between the IDT and qubit. Several further frequency dips are then observed, spaced by approximately $\unit[430]{ns}$, again consistent with the SAW time-of-flight from qubit to one Bragg mirror and back again. The qubit frequency is observed to decay back to its undisturbed value over a timescale of $\approx 3~\rm{\mu s}$, similar to the phonon lifetime of the SAW cavity. The multiple reflections between qubit, IDTs, and Bragg mirrors likely serve to smooth out the response beyond a time delay of around $1~\rm{\mu s}$. As well as demonstrating the unique slow propagation of SAWs in a quantum device, this experiment also serves to further prove that the Stark shifts that we observe are indeed due to the acoustic field of the SAW mode, rather than a crosstalk of the electromagnetic signal applied to the IDT directly to the qubit.

The prototype quantum acoustic device that we have presented here may be significantly improved, opening up the possibility of using cavity-trapped SAWs for quantum memories, time delays, and quantum signal filtering applications. In particular, we have used a relatively weak piezoelectric substrate for our experiment, nevertheless achieving a qubit-SAWR coupling strength of 5.7 MHz. Stronger piezoelectrics such as lithium niobate or zinc oxide could dramatically increase this coupling strength. This could have the additional benefit of enabling the qubit coherence to be improved, as the electric field of the qubit could be designed to only partially rather than fully reside in the piezoelectric substrate (which in the present case likely limits coherence due to undesired bulk acoustic emission). The $10^5$-times reduced speed of travel of SAWs compared to electromagnetic signals also makes our device a miniaturised mechanical implementation of traditional cavity QED, and an ideal engineered platform to push the boundaries of cavity QED physics, opening up the possibility to explore, for instance, strong coupling multimode cavity QED with mechanical devices.

This work has received funding from the UK Engineering and Physical Sciences Research Council under Grant Nos. EP/J001821/1 and EP/J013501/1 and from the Japan Society for the Promotion of Science.

\bibliographystyle{unsrt}
\bibliographystyle{apsrev4-1}
\bibliography{Manenti150521} 

\clearpage


\pagebreak
\widetext
\begin{center}
\textbf{\large Supplementary Information}
\end{center}
\setcounter{equation}{0}
\setcounter{figure}{0}
\setcounter{table}{0}
\setcounter{page}{1}
\makeatletter
\renewcommand{\theequation}{S\arabic{equation}}
\renewcommand{\thefigure}{S\arabic{figure}}
\renewcommand{\thetable}{S\arabic{table}}
\renewcommand{\bibnumfmt}[1]{[#1]}
\renewcommand{\citenumfont}[1]{#1}

\section{I. Device parameters}
\label{section1}

\begin{figure}[b]
\centering
\includegraphics[width=0.95\linewidth]{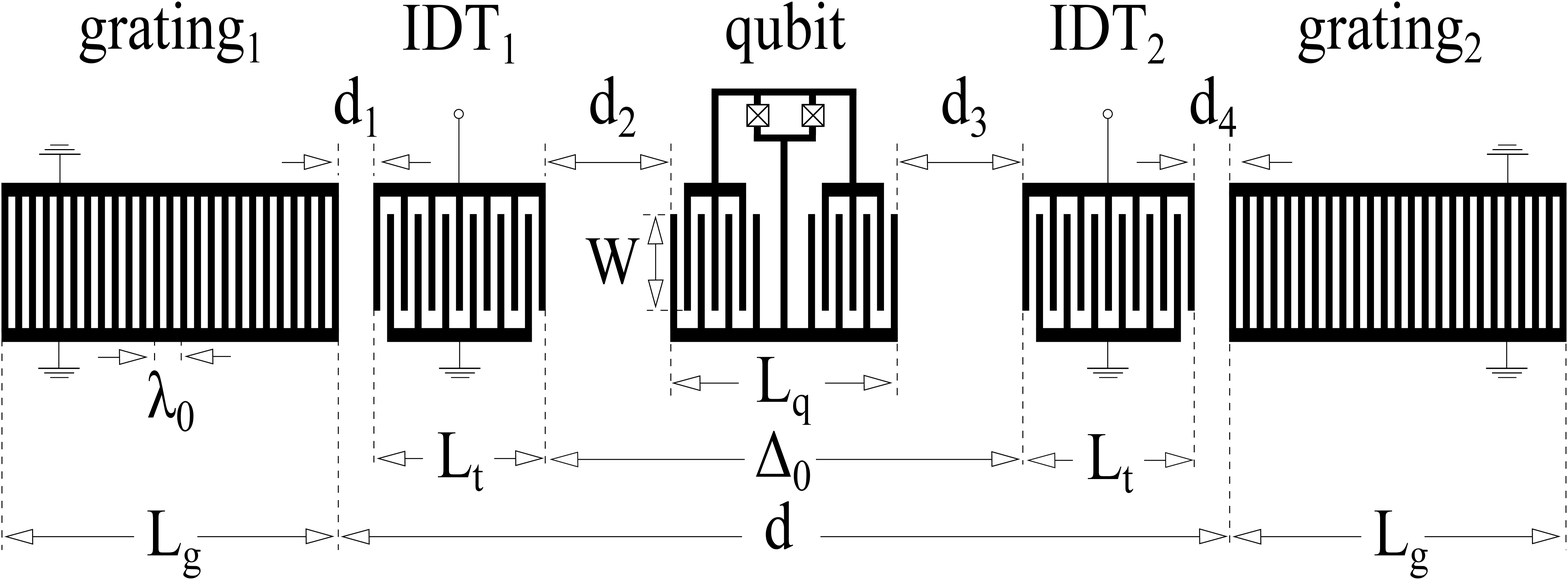}
\caption{\label{fig:supp1} \textbf{$\!\!$Figure S1 $|$ Schematic of a transmon qubit embedded in a 2-port SAW resonator.} The 2-port SAW resonator consists of two IDTs and two gratings. The tuneable qubit is formed by a SQUID shunted by an interdigitated capacitance. Due to limited space, the device represented in this figure is not in scale and the number of fingers in the qubit, IDTs and gratings has been reduced with respect to the real device (see text for further details).}
\end{figure}
In Fig.~\ref{fig:supp1}, we present a schematic of a tuneable transmon embedded in a 2-port SAW resonator. The acoustic resonator consists of two interdigitated transducers (IDTs) and two gratings. Each IDT is an interdigitated capacitance with $N_\textrm{t}=51$ fingers. The periodicity of the transducers is $\lambda_0 = \unit[6]{\mu m}$ and their horizontal length is $L_\textrm{t}=(2 N_\textrm{t}-1)\lambda_0/4=\unit[151.5]{\mu m}$. One bus bar of the IDT is grounded whereas the opposite one is connected to a waveguide. As regards the gratings, they consist of $N_\textrm{g}=400$ shorted metallic strips connected to ground with a periodicity of $\lambda_0/2$. The horizontal length of each grating is $L_\textrm{g}=(2 N_\textrm{g}-1)\lambda_0/4=\unit[1198.5]{\mu m}$. In the remainder, we will call the left and right transducers $\textrm{IDT}_1$ and $\textrm{IDT}_2$ respectively. The same nomenclature is adopted for the gratings. 
 
For optimal reflection, the minimal distance between $\textrm{IDT}_1$ and $\textrm{grating}_1$ must satisfy the  relation $d_1=(n_1/2 - 1/4) \lambda_0 - \lambda_0 /8$, where $n_1 \in \mathbb{N}^+$. To minimise the cavity length we opted for $n_1=3$, leading to $d_1 = \unit[6.75]{\mu m}$ (a lower value of $n_1$ might have caused some fabrication issues). The distance $d_4$ between $\textrm{IDT}_2$ and $\textrm{grating}_2$ is the same as $d_1$. Finally, the minimal distance between the two IDTs satisfies the relation $\Delta_0 = (n_2 - 3/4) \lambda_0$ where $n_2 \in \mathbb{N}^+$. The value of $n_2$ can be chosen based on some considerations. From previous experiments\textcolor{blue}{$^1$}, it has been noted that the internal quality factor of a SAW resonator increases as the distance between the two gratings increases. It has also been observed that the number of longitudinal modes supported by the acoustic cavity increases with increasing distance between the two gratings. Lastly, the coupling between the qubit and the SAWR is inversely proportional to the square root of the acoustic cavity area (Supplementary Information, section II). For a better coherence, we aimed at having an internal quality factor as high as possible; at the same time, we wanted to maximise the acoustic coupling and we wanted our cavity to support a small number of longitudinal modes. For these reasons, we have set $n_2$ to the reasonable value of 158 and therefore $\Delta_\textrm{0} = \unit[943.5]{\mu m}$. As explained in the main text, our acoustic cavity supports three longitudinal modes separated by about $\unit[1]{MHz}$.

The tuneable transmon is placed in the middle of the acoustic cavity. It consists of a $ \unit[4 \times 4.57]{\mu m^2}\,$ SQUID shunted by an interdigitated capacitance. The capacitance of the qubit resembles a transducer itself: it consists of $N_\textrm{q}=30$ fingers with one additional central finger connecting the two electrodes of the capacitance (see Fig.~\ref{fig:supp1}). The interdigitated structure of the qubit capacitance has the same periodicity as the IDTs. Unlike these acoustic components, the qubit is connected neither to ground nor to any other waveguide (it is only capacitively coupled to ground and to a CPWR not shown in the figure). 

The length of the fingers in the IDTs and qubit is denoted by $W$. A large value of $W$ would considerably decrease diffraction losses\textcolor{blue}{$^1$}; at the same time, a small value of $W$ would decrease the acoustic cavity area and increase the acoustic coupling (Supplementary Information, section II). A tradeoff between these two limits lead us to $W=11.66 \lambda_0 = \unit[70]{\mu m}$. Note that at low temperatures, the device undergoes a non-isotropic contraction and the distances between, as well as the dimensions of the acoustic components may vary on the order of tens of nanometers.

\begin{figure}[t]
\centering
\includegraphics[width=0.95\linewidth]{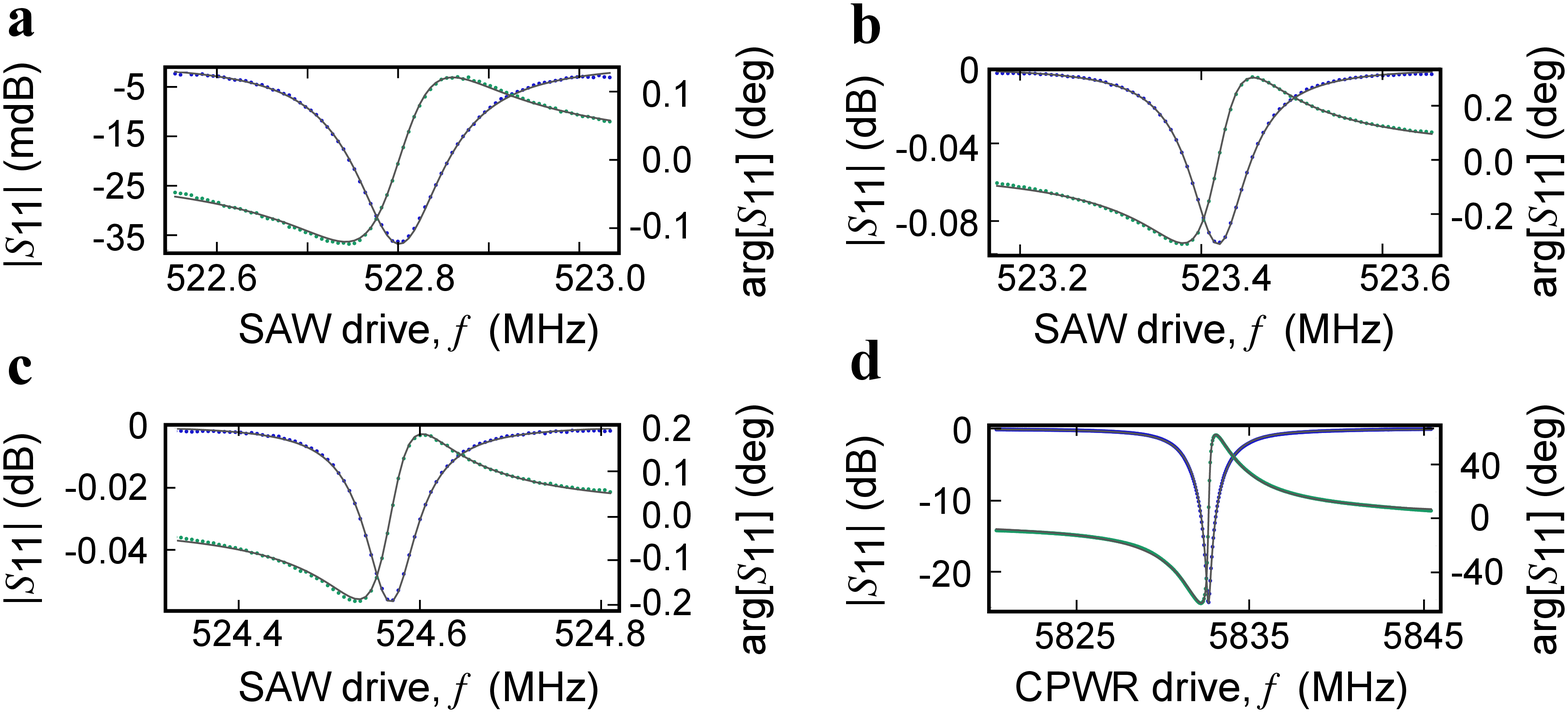}
\caption{\label{fig:supp2} \textbf{$\!\!$Figure S2 $|$ Reflection coefficient of the 2-port SAWR and CPWR.} \textbf{a-c},
Magnitude (blue) and phase (green) of the measured reflection coefficient $S_{11}(f)$ of the longitudinal modes $f_\textrm{m1}$, $f_\textrm{m2}$ and $f_\textrm{m3}$. Solid lines are a fit to equation~\eqref{eq:s11}.
\textbf{d},
Magnitude (blue) and phase (green) of the measured reflection coefficient $S_{11}(f)$ of the CPWR mode $f_\textrm{r}$. Solid lines are a fit to equation~\eqref{eq:s11}. A background due to the measurement setup has been subtracted in all these frequency responses.}
\end{figure}

By applying an oscillating voltage to one $\textrm{IDT}$, it is possible to generate a surface acoustic wave. The frequency of the wave is given by the simple formula $f= v_\textrm{e} /\lambda$, where $v_\textrm{e}$ is an effective speed of sound. An optimal transduction is achieved when $\lambda = \lambda_0 $, which leads to the central resonant frequency of the device $f_0 = v_\textrm{e} /\lambda_0$.  Assuming that $f_\textrm{m2}$ is the central resonance, we can extract an effective speed of sound of $v_\textrm{e}= \lambda_0 \times f_\textrm{m2} = \unit[3140.6]{m/s}$. The acoustic wave is eventually collected by the second transducer generating a potential difference on its electrodes. In our setup, we used a vector network analyzer (VNA) to acquire the transmitted signal from $\textrm{IDT}_1$ to $\textrm{IDT}_2$; the results of this measurement are shown in Fig.~\ref{fig:2}\textcolor{blue}{a} of the main text. The SAWR can also be measured in reflection. In Fig.~\ref{fig:supp2}\textcolor{blue}{a-c} we present the measured reflection coefficient $S_{11}(f)$ of the 2-port SAW resonator around the three longitudinal modes $f_\textrm{m1}$, $f_\textrm{m2}$ and $f_\textrm{m3}$. Close to resonance, the SAWR can be modelled with an $RLC$ equivalent circuit. According to this model, the analytical expression of the reflection coefficient takes the form:
\begin{equation}
S_{11}\left(f \right)=\frac{\left(Q_\mathrm{m,e} - Q_\mathrm{m,i} \right)/Q_\mathrm{m,e}+2iQ_\mathrm{m,i}\left(f-f_\textrm{m}\right)/f}{\left(Q_\mathrm{m,e} + Q_\mathrm{m,i} \right)/Q_\mathrm{m,e}+2iQ_\mathrm{m,i}\left(f-f_\textrm{m}\right)/f}.\label{eq:s11}
\end{equation}
Here $Q_\mathrm{m,i}$ is the internal quality factor of the mechanical mode and $Q_\mathrm{m,e}$ is the external quality factor due to the presence of the IDT and measurement port. From a fit to the experimental data, we find $Q_\mathrm{m1,i}=4830$, $Q_\mathrm{m2,i}=7020$, $Q_\mathrm{m3,i}=7600$, $Q_\mathrm{m1,e}=2.34\times 10^6$, $Q_\mathrm{m2,e}=1.33\times 10^6$, and $Q_\mathrm{m3,e}=2.33\times 10^6$. In terms of loss rate, these values can be expressed as $\kappa_\textrm{m1}=\unit[108]{kHz}$, $\kappa_\textrm{m2}=\unit[75]{kHz}$ and $\kappa_\textrm{m3}=\unit[69]{kHz}$.

For completeness, we also report the measured reflection coefficient of the CPWR in Fig.~\ref{fig:supp2}\textcolor{blue}{d}. Its resonant frequency depends on its length $L_\textrm{r}$ and the effective dielectric constant of the substrate $\varepsilon_\textrm{eff}$ in the following way: $f_\textrm{r}=c/2 L_\textrm{r}\sqrt{\varepsilon_\textrm{eff}}$, where $L_\textrm{r} = \unit[14100]{\mu m}$, and $\varepsilon_\textrm{eff} \approx 3.3$ for quartz. Finally, by fitting the CPWR response in the frequency domain, we obtain $Q_\mathrm{m2,i}=3240$ and $Q_\mathrm{m3,e}=3630$. In terms of loss rate, these values can be expressed as $\kappa_\textrm{r}=\unit[3.41]{MHz}$.

\begin{table}
\begin{tabular}{llll}
\hline \\[-8pt]
Qubit & Coulomb energy & $E_{\textrm{C}}=\unit[0.310]{GHz}$ & \\[3pt]
 & Maximum Josephson energy & $ E_{\textrm{J0}} =\unit[10.704]{GHz}$ & \\[3pt]
 & $ E_{\textrm{J0}} / E_{\textrm{C}}$ & $34.5$ & \\[3pt]
 & Relaxation time & $T_{1}=\unit[46]{ns}$ (at $f_\textrm{q}=\unit[2.6]{GHz}$)& \\[3pt]
 & Dephasing time & $T_{2}=\unit[67]{ns}$ (at $f_\textrm{q}=\unit[2.6]{GHz}$) & \\[3pt]
 & Qubit frequency & $f_\textrm{q}(\Phi)=\left(\sqrt{8E_{\textrm{C}}E_{\textrm{J}0}\cos\left|\pi \Phi / \Phi_0 \right|}-E_{\textrm{C}}\right)/\hbar$ & \\[3pt]
 \hline \\[-9pt]
 & & \\[-8pt]
CPWR & Resonant frequency and linewidth & $f_{\textrm{r}}=\unit[5.83]{GHz} \qquad \kappa_{\textrm{r}}= \unit[3.41]{MHz} $ &  \\[3pt]
 & Internal and external quality factor & $Q_{\textrm{r,i}}=3240 \qquad Q_{\textrm{r,e}}=3630$ &  \\[3pt]
 & Qubit - CPWR coupling strength & $g=\unit[69]{MHz} $ &  \\[3pt]
\hline \\[-8pt]
SAWR & Resonant frequencies and linewidths & $f_{\textrm{m1}}=\unit[522.825]{MHz}\qquad \kappa_{\textrm{m1}}= \unit[108]{kHz} $ & \\[3pt]
 & (in reflection) & $f_{\textrm{m2}}=\unit[523.426]{MHz}\qquad \kappa_{\textrm{m2}}= \unit[75]{kHz}$ & \\[3pt]
 &  & $f_{\textrm{m3}}=\unit[524.575]{MHz}\qquad  \kappa_{\textrm{m3}}= \unit[69]{kHz}$& \\[3pt]
 & Internal/external quality factors & $Q_{\textrm{m1,i}}= 4830 \qquad Q_{\textrm{m1,e}}=2.34\times 10^6$ & \\[3pt]
 & (in reflection)  & $Q_{\textrm{m2,i}}=7020 \qquad Q_{\textrm{m2,e}}=1.33\times 10^6$ &\\[3pt]
 &  & $Q_{\textrm{m3,i}}=7600 \qquad Q_{\textrm{m3,e}}=2.33\times 10^6$ &\\[3pt]
 & Periodicity & $\lambda_0=\unit[6]{\mu m}$ & \\[3pt]
 & Effective speed of sound & $v_{\textrm{e}}=\unit[3140.6]{m/s}$ & \\[3pt]
 & Fingers in each IDT & $N_{\textrm{t}}=51$ & \\[3pt]
 & Fingers in each grating & $N_{\textrm{g}}=400$ & \\[3pt]
 & Fingers in the qubit capacitance & $N_{\textrm{q}}=31$ & \\[3pt]
 & Length of each finger & $W=\unit[70]{\mu m}=11.66 \lambda_0$ & \\[3pt]
 & Distance between grating and IDT & $d_1 = \left.(n_1/2 - 1/4) \lambda_0 - \lambda_0 /8 \right|_{n_1=3}= \unit[6.75]{\mu m} $ & \\[3pt]
 & Distance between gratings & $d= \left.2 d_1 + 2 \left(2 N_t-1\right)\lambda_0/4 + (-3/4 + n_2)\lambda_0 \right|_{n_2=158} $ =& \\[3pt]
 & & $\phantom{d}= \unit[1260]{\mu m}=  v_{\textrm{e}} \times \unit[401]{ns}$ & \\[3pt]
 & Distance between centre of IDTs & $d_{\textrm{IDT}}= \unit[1100]{\mu m} $ & \\[3pt]
 & Cavity length & $L_{\textrm{c}}=\unit[1365]{\mu m} $& \\[3pt]
 & Cavity area & $A =  W\times L_\textrm{c}  = \unit[95900]{\mu m^2} $ & \\[3pt]
 & Qubit - SAWR coupling strength & $\lambda_\textrm{m2}(f_\textrm{q}=\unit[2.5]{GHz})=\unit[5.7]{MHz} $ &  \\[3pt]
\end{tabular}
\caption{\label{tab:supp1} \textbf{$\!\!$Table S1 $|$ Device parameters.}}
\end{table}

When the wave bounces against the mirrors, it slightly penetrates into this regular array of metallic fingers. The distance that the wave travels into this periodic structure is called penetration depth $L_\textrm{p}$.  The longitudinal cavity length is given by the sum of the distance between the two gratings and the penetration depth into them: $L_\textrm{c}=d + 2 L_\textrm{p}$. From the frequency difference between $f_\textrm{m3}$ and $f_\textrm{m2}$, we can extract the cavity length: $L_\textrm{c}=  v_\textrm{e}/2|f_\textrm{m3}-f_\textrm{m2}|=\unit[1370]{\mu m}$. As explained in the main text, this value is in agreement with measurements performed in the time domain. From the value of the cavity length, we can derive the penetration depth $L_\textrm{p} = \unit[55]{\mu m} $. The reflectivity $|r_s|$ of each finger in the grating can be easily obtained from the relation $4 |r_s| L_\textrm{p} = \lambda_0 \tanh (|r_s| N_\textrm{g})$, whence $r_s = |0.0273|$.
All of the parameters presented so far are listed in Table~\ref{tab:supp1}.

\begin{figure}[t]
\centering
\includegraphics[width=0.95\linewidth]{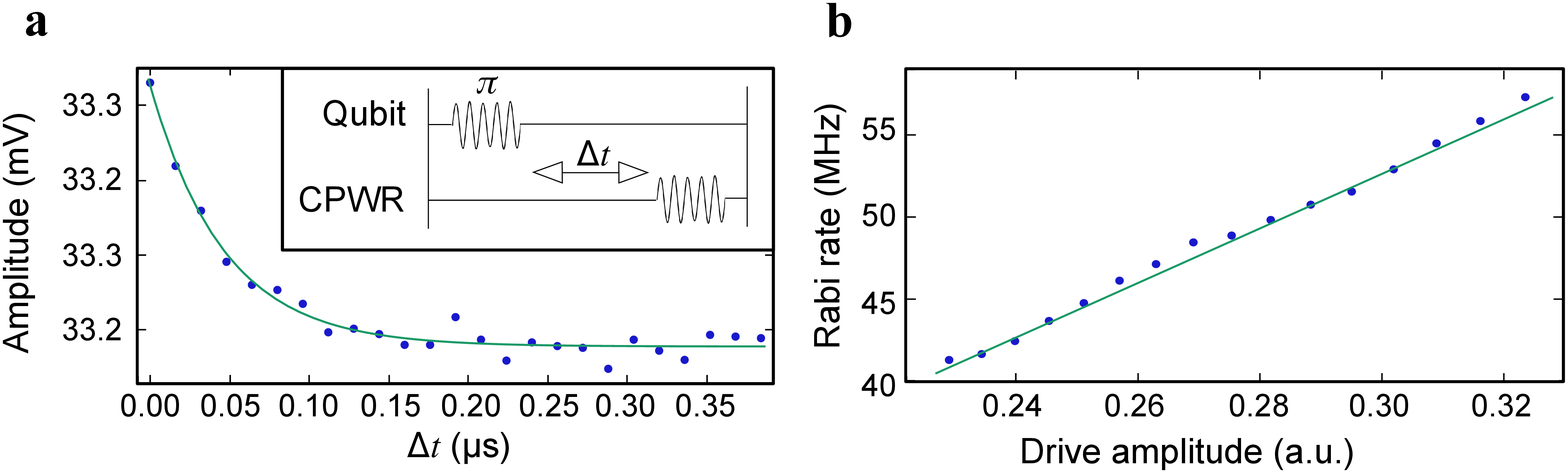}
\caption{\label{fig:supp3-1} \textbf{$\!\!$Figure S3 $|$ Qubit coherence.} \textbf{a}, Inversion recovery experiment to extract the qubit decay time $T_1$. The solid green curve based on equation~\eqref{eq:exponentialmodel} fits well the data points (blue dots). \textbf{b}, Rabi rate for different values of drive amplitude (blue points). As expected, the data points fit well to a linear dependence (green solid line).}
\end{figure}

We conclude this section presenting some observations concerning the coherence time of our qubit. In order to extract the decay time $T_1$ of our transmon, we performed an inversion recovery experiment. The pulse scheme used to extract $T_1$ is shown in the inset of Fig.~\ref{fig:supp3-1}\textcolor{blue}{a}: a $\pi$ pulse is applied to the qubit followed by a readout pulse at the CPWR frequency. Figure~\ref{fig:supp3-1}\textcolor{blue}{a} shows the exponential decay of the qubit population as a function of the delay between the two pulses. The data points fit well to the exponential model:
\begin{equation}
P_\textrm{e} (t) = \frac{1+\langle\hat{\sigma}_z (t)\rangle}{2}= A + B \exp{(-\Delta t/T_1)}, \label{eq:exponentialmodel}
\end{equation}
where $A=\unit[32.878]{mV}$ is an offset, $B=\unit[0.3496]{mV}$ is a scaling factor and $T_1 =\unit[46]{ns}$ is the decay time of the qubit (this experiment has been performed with the qubit frequency fixed at $f_\textrm{q}=\unit[2.9]{GHz}$). We have also performed Rabi oscillations of the qubit using the CPWR as readout. Figure~\ref{fig:supp3-1}\textcolor{blue}{b} shows the frequency of Rabi oscillations as a function of drive amplitude: as expected, there is a linear dependence between these two variables.

\section{II. Coupling strength between a charge qubit and a SAW cavity}
When a surface perturbation on a piezoelectric crystal travels through an interdigitated capacitor with the same periodicity as the incoming wave, the capacitor develops an oscillating voltage on its electrodes. This phenomenon can be exploited to couple a surface acoustic wave to a transmon with a suitably shaped capacitance. The coupling strength between a transmon and a SAW cavity can be calculated by considering the charge $q$ and the potential difference $V_0$ generated by a single phonon on the electrodes of the transmon. Let us first derive the potential difference $V_0$. The zero-point mechanical motion associated to a single phonon inside a SAW cavity is:
\begin{equation}
U_0 = \sqrt{\frac{\hbar}{2  \rho A_\textrm{c} v_\textrm{e}}}, \label{eq:U0}  
\end{equation}
where $A = W \times L_\textrm{c} = \unit[95900]{\mu m^2} $ is the area of the acoustic cavity and $\rho = \unit[2647]{kg / m^3}$ is the quartz mass density. From the zero-point mechanical motion, we can easily derive the value of the zero-point electric potential:
\begin{equation}
\phi_0 \approx \frac{e_\textrm{pz}}{\varepsilon}U_0= \frac{e_\textrm{pz}}{\varepsilon}\sqrt{\frac{\hbar}{2  \rho A_\textrm{c} v_\textrm{e}}}, \label{eq:phi0}
\end{equation}
where $\varepsilon$ 
is the permittivity of the substrate and $e_\textrm{pz}$ is a component of the quartz piezoelectric tensor which depends on the propagation direction (for ST-X quartz\textcolor{blue}{$^2$}, $e_\textrm{pz}/\varepsilon\approx\unit[2.0]{V/nm}$).
As mentioned earlier, the transmon capacitance responds in a more effective way to waves sharing the same periodicity of its structure. Hence, the electric potential $\phi_0$ has to be scaled according to the following normalised array factor\textcolor{blue}{$^2$}:
\begin{equation}
A(f)=  \left|\frac{\sin \left[N_\textrm{q} \pi (f - f_0)/2f_0\right]}{N_\textrm{q} \pi (f - f_0)/2f_0}\right|. \label{eq:array}
\end{equation}
Note that for $f=f_0$, $A(f) =  1$. The potential difference is thus $V_0 = \phi_0 A(f)$. As regards the charge generated by the surface acoustic wave on the transmon electrodes, its value is given by:
\begin{equation*}
\hat{q} =  2e \beta  \hat{n}  \label{eq:charge}
\end{equation*}
where $2e$ is the charge of a Cooper pair, $\beta$ is a prefactor, $|j\rangle$ is an eigenstate of the transmon and $\hat{n}$ is a quantum operator indicating the number of Cooper pairs in excess (or deficit) on the superconducting island. To calculate the coupling in the transmon eigenbasis, we need the matrix element\textcolor{blue}{$^3$}:
\begin{equation*}
\langle j+1 | \hat{q} | j \rangle =  2e \beta \langle j+1 | \hat{n} |j \rangle \approx 2e \beta \sqrt\frac{j+1}{2} \left(\frac{E_\textrm{J} (\phi)}{8 E_\textrm{C}}\right)^{1/4}.
\end{equation*}
For the first two levels of the transmon, we have $\langle 1 | \hat{q} |0 \rangle \approx  e \beta\left(E_\textrm{J} (\phi)/2 E_\textrm{C}\right)^{1/4}$.
It remains to calculate the prefactor $\beta$. This parameter originates from the fact that not all of the charge generated by the surface acoustic wave will be localised on the transmon capacitance: part of it will be distributed on the gate capacitance $C_2$, on the junction capacitance $C_\textrm{J}$ and strain capacitances $C_s$ to ground planes and other metallic components of the chip. Hence:
\begin{equation*}
\beta = \frac{C_\textrm{q}}{C_2 + C_s + C_\textrm{J}+C_\textrm{q}}=\frac{C_\textrm{q}}{C_\Sigma}.
\end{equation*}
where $C_\textrm{q}=W N_\textrm{q} \varepsilon /2$ is the capacitance of the qubit interdigitated structure.
The coupling strength between a transmon and a SAW cavity can be written as follows:
\begin{eqnarray*}
\lambda (\phi , f) &=& \frac{\langle 1| \hat{q} |0\rangle \,V_0}{h}  \approx \frac{2e}{h}  \beta \langle 1 | \hat{n} |0 \rangle \phi_0 A(f)=\\
&=& \frac{e\beta}{h}   \; \left(\frac{E_\textrm{J} (\phi)}{2 E_\textrm{C}}\right)^{1/4}  \phi_0 A(f),
\end{eqnarray*}
where the value of $E_\textrm{J} (\phi)$ depends on the qubit frequency in the following way $E_\textrm{J} (\phi)= [hf_\textrm{q}(\phi) + E_\textrm{C}]^2/8 E_\textrm{C}$.
Substituting values related to our device and assuming that the frequency of the qubit is $f_\textrm{q}=\unit[2.52]{GHz}$  and approximating $\beta \approx 1$ and $A(f) \approx 1$, we have:
\begin{eqnarray*}
\lambda &\approx&  \frac{e}{h}    \left(\frac{\unit[3.4]{[GHz]} }{2.0 \times \unit[0.31]{[GHz]}}\right)^{1/4} \unit[2]{\left[\frac{nV}{m}\right]} \sqrt{\frac{\hbar}{2\times\unit[ 2647]{[kg /m^{3}]} \times \unit[95900]{[\mu m^2]} \times\unit[3140.6]{[m/s]} }}  =\\
&=& \unit[6.0]{MHz}.
\end{eqnarray*}
This value agrees well with the experimental value extracted from our measurements.

\section{III. Cryogenic setup and fabrication procedure}

\begin{figure}[t]
\centering
\includegraphics[width=0.95\linewidth]{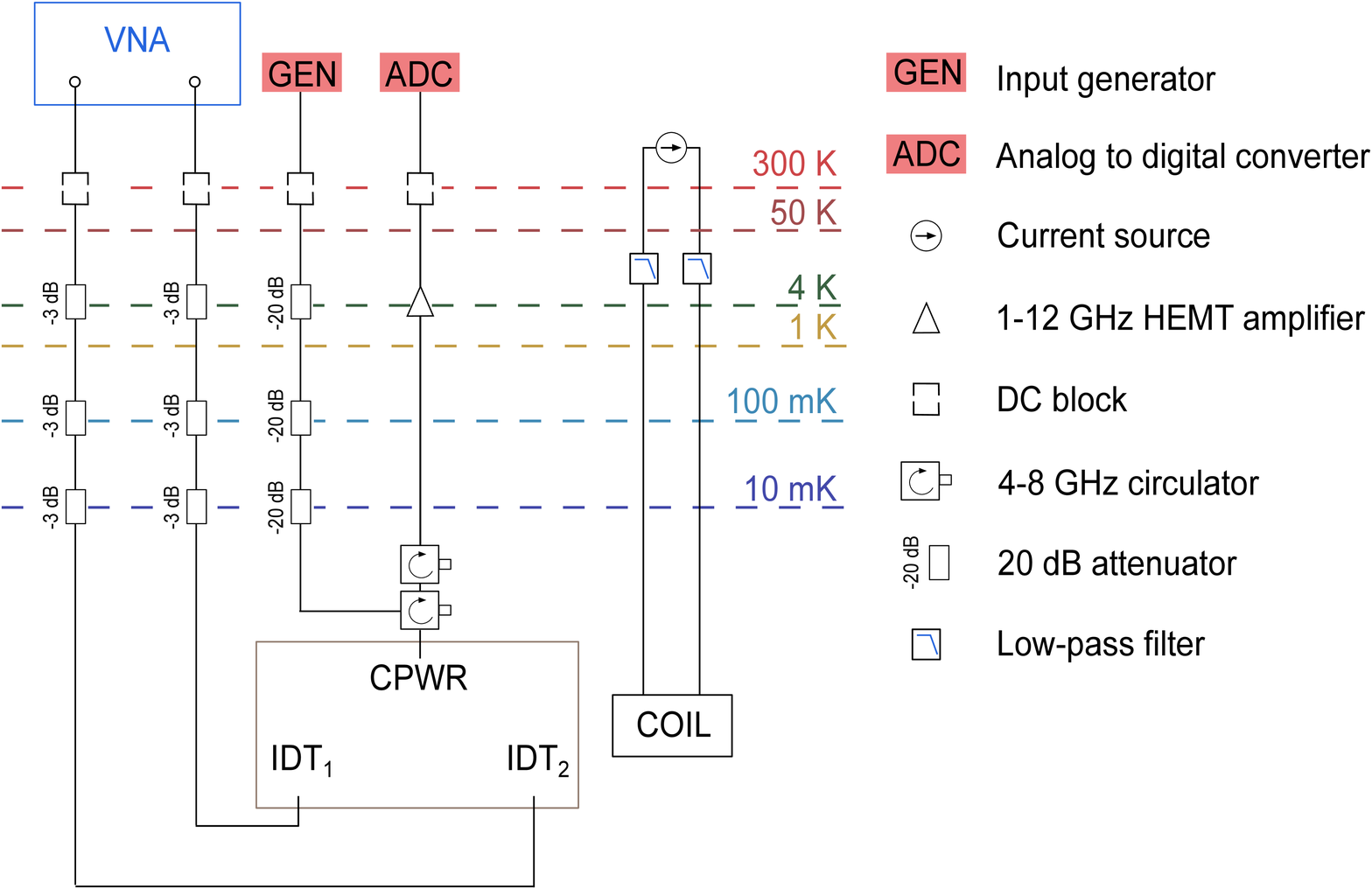}
\caption{\label{fig:supp4} \textbf{$\!\!$Figure S4 $|$ Cryogenic setup.}}
\end{figure}

The device presented in this work has been characterised at cryogenic temperatures. The microchip has been bonded on a circuit board, placed inside a home-made oxygen-free copper sample holder and mechanically anchored to the $\unit[10]{mK}$ plate of a dilution refrigerator (Triton200, Oxford Instruments). The microwave line connecting the CPWR to the external instrumentation is highly attenuated by means of three $\unit[-20]{dBm}$ attenuators. The measured  attenuation of this line is $\unit[-67]{dB}$.  The reflected signal coming from the CPWR passes through two \unit[4-8]{GHz} circulators and reaches a 1-12$\,$GHz HEMT cold amplifier (see Fig.~\ref{fig:supp4}). The signal is then downconverted and acquired using an analog-to-digital converter at room temperature. The lines connecting $\textrm{IDT}_1$ and $\textrm{IDT}_2$ to the VNA have an estimated overall attenuation of $\unit[-16]{dB}$.

We conclude this section presenting the fabrication procedure of our device. First of all, the ground planes, alignment marks and waveguides have been patterned with standard photolithographic techniques. These $\unit[100]{nm}$ thick metallic structures have been deposited with a home-made electron-beam evaporator. The SAWR and the qubit have been patterned together in a second electron-beam lithography step. The $\unit[200\times200]{nm^2}$ junctions have been fabricated with the usual Dolan bridge technique and with a double angle evaporation in the following way: firstly $\unit[30]{nm}$ of aluminium have been deposited, followed by an in-situ oxidation step and a second deposition of $\unit[60]{nm}$ of aluminium. The overall height of the qubit and SAWR is therefore $\unit[90]{nm}$.

\begin{center}
\rule{0.3\textwidth}{.4pt}
\end{center}
[1] \small{R. Manenti, M. J. Peterer, A. Nersisyan, E. B. Magnusson, A. Patterson, and P. J. Leek, \href{http://link.aps.org/doi/10.1103/PhysRevB.93.041411}{\textcolor{blue}{Phys. Rev. B \textbf{93}, 041411 (2016)}}.} \newline
[2] \small{D. Morgan, \href{https://www.elsevier.com/books/surface-acoustic-wave-filters/morgan/978-0-12-372537-0}{\textcolor{blue}{\textit{Surface Acoustic Wave Filters}}} (Academic Press, Amsterdam, 2007).}\newline
[3] \small{J. Koch, T. M. Yu, J. Gambetta, A. A. Houck, D. I. Schuster, J. Majer, A. Blais, M. H. Devoret, S. M. Girvin, and R. J.}\newline
\indent $\,\!$ \small{Schoelkopf, \href{http://journals.aps.org/pra/abstract/10.1103/PhysRevA.76.042319}{\textcolor{blue}{Phys. Rev. A \textbf{76}, 042319 (2007)}}.}\newline

\end{document}